\documentclass[a4paper,11pt]{article}
\usepackage{pos}

\title{\huge Unification of Gauge Symmetries\\ ... including their breaking}
\ShortTitle{Unification of Gauge Symmetries... including their breaking}

\author[a]{Andrei Angelescu}
\author[a]{Andreas Bally}
\author[b]{Simone Blasi}
\author*[a]{Florian Goertz}

\affiliation[a]{Max-Planck-Institut f{\"u}r Kernphysik,\\ Saupfercheckweg 1, 69117 Heidelberg, Germany}

\affiliation[b]{Theoretische Natuurkunde \& IIHE/ELEM, Vrije Universiteit Brussel, and International Solvay Institutes,\\ Pleinlaan 2, B-1050 Brussels, Belgium}

\emailAdd{andrei.angelescu@mpi-hd.mpg.de}
\emailAdd{andreas.bally@mpi-hd.mpg.de}
\emailAdd{simone.blasi@vub.be}
\emailAdd{florian.goertz@mpi-hd.mpg.de}

\abstract{In this talk, we present a minimal viable scenario that unifies the gauge symmetries of the Standard Model (SM) and their breaking sector. Our Gauge-Higgs Grand Unification setup employs 5D warped space with a $SU(6)$ bulk gauge field that includes both a $SU(5)$ grand unified theory (GUT) and a Higgs sector as a scalar component of the 5D vector field, solving the hierarchy problem. By appropriately breaking the gauge symmetry on the boundaries of the extra dimension the issue of light exotic new states, appearing generically in such models, is eliminated and the SM fermion spectrum is naturally reproduced. The Higgs potential is computed at one-loop, finding straightforward solutions with a realistic $m_h = 125$\,GeV. The problem of proton decay is addressed by showing that baryon number is a hidden symmetry of the model. The presence of a scalar leptoquark and a scalar singlet is highlighted, which might play a role in solving further problems of the SM, allowing for example for electroweak baryogenesis. Finally, the $X$ and $Y$ gauge bosons from $SU(5)$ GUTs are found at collider accessible masses, opening a window to the unification structure at low energies.}

\FullConference{%
  *** The European Physical Society Conference on High Energy Physics (EPS-HEP2021), ***\\
  *** 26-30 July 2021 ***\\
  *** Online conference, jointly organized by Universität Hamburg and the research center DESY ***
}

\begin{document}
\maketitle

\section{Introduction}

It is a long standing dream of fundamental physics to trace back the basic interactions of nature to a single symmetry group of a `grand unified theory' (GUT)~\cite{PhysRevLett.32.438,PhysRevD.10.275}. The road to its realization, however, comes with various challenges, some of which include 
fast proton decay and the doublet-triplet splitting problem, i.e., the issue of the Higgs doublet being generically degenerate with its heavy color-triplet partner.
Beyond that, GUTs make the so-called hierarchy problem (HP) manifest, denoting the fact that the electroweak (EW) scale receives large quantum corrections that pull it towards the (too large) scale of grand unification.

Leaving these challenges aside, it would be fascinating for its own sake to push forward the concept of unification by not only unifying the fundamental gauge symmetries, but also including the sector that spontaneously breaks them -- leading to a non-linearly realized EW symmetry with massive gauge bosons at low energies -- in a single entity. In fact, in scenarios of Gauge-Higgs Unification (GHU)~\cite{Manton:1979kb,Fairlie:1979at,Hosotani:1983vn,Hosotani:1983xw} the Higgs scalar is realized as the fifth component of a 5D gauge~field~$A_5^{\!(0)}$\,(in a (warped) extra dimension~\cite{Randall:1999ee,Contino:2003ve}), see~\cite{Agashe:2004rs,Medina:2007hz,Hosotani:2008tx,Funatsu:2013ni,Funatsu:2014fda} for a realistic application to the EW theory.

Here, we discuss a new economical setup \cite{Angelescu:2021nbp} that uses this GHU idea to unify a $SU(5)$ GUT gauge group with the Higgs sector of spontaneous symmetry breaking in a {\it single} 5D field -- a framework which is called Gauge-Higgs Grand Unified Theory (GHGUT). Beyond this unification, the HP is solved since its embedding into a gauge field, together with the extra-dimensional geometry, protects the Higgs boson from large corrections to its mass.\footnote{Taking a dual 4D perspective~\cite{ArkaniHamed:2000ds}, the Higgs boson corresponds to a composite pseudo Nambu-Goldstone Boson (pNGB) of a spontaneously broken global symmetry, with the compositeness scale cutting off corrections to $m_h$,~see~\cite{Angelescu:2021nbp}.}
% addressing the remaining little hierarchy between the Higgs mass and the compositeness scale. 

From a different point of view, such models can be seen as a natural extension of GHU~and~the pNGB-Higgs, since they employ a set of generators of an enlarged symmetry, characteristic of these frameworks (see \cite{Bellazzini:2014yua}), to realize a GUT. 
Finally, the construction~\cite{Angelescu:2021nbp} also solves the other mentioned issues of GUTs, i.e., proton decay and doublet-triplet degeneracy and is in principle accessible at colliders. All of this is achieved without the emergence of problematic ultra-light~exotic states and with a Higgs mass residing easily at $m_h=125\,$ GeV, while keeping the concept~of~minimality -- overcoming difficulties of similar proposals \cite{Hosotani:2015hoa,Furui:2016owe,Hosotani:2016njs,Hosotani:2017edv,Hosotani:2017hmu,Hall:2001zb,Burdman:2002se,Haba:2004qf,Lim:2007jv,Maru:2019lit,Maru:2019bjr}, which for example require more dimensions or large (ad-hoc) extensions of the matter sectors to become viable. The model discussed here thus follows closely the spirit of the canonical composite Higgs~\cite{Contino:2003ve,Agashe:2004rs,Contino:2006qr} (for reviews, see~\cite{Contino:2010rs,Azatov:2012qz,Bellazzini:2014yua,Panico:2015jxa,Goertz:2018dyw}).

%%%%%%%%%%%%%%%%%%%%%%%%%%%%%%%%%%%%%%%%
\section{Model}
\label{sec:setup}
We consider a $G=SU(6)$ bulk gauge symmetry in a slice of warped AdS$_5$ space with metric
\begin{equation}
 ds^2 = \left(R/z\right)^2 
   \left( \eta_{\mu\nu}\,dx^\mu dx^\nu - dz^2 \right)\,,
\end{equation}
where $z\in[R,R']$, and $R \sim 1/M_{\rm PL}$ ($R^\prime \sim 1/\rm TeV$) is the position of the UV (IR) brane, addressing the HP. To arrive at a low-energy theory that resembles the Standard Model (SM), the $SU(6)$ bulk symmetry is broken to subgroups $H_0$ and $H_1$ on the UV and the IR branes, respectively, by gauge boundary conditions (BCs), inducing~\cite{Angelescu:2021nbp}
\begin{equation}
\label{eq:breakingmod}
    \begin{split}
SU(6) & \to SU(5) \equiv H_0,\\
SU(6) & \to SU(2)_L \times SU(3)_c \times U(1)_Y \equiv H_1\,.
    \end{split}
\end{equation}
We see that the remaining unbroken gauge group is exactly  $G_{\rm SM} = H_1 = H_0 \cap H_1 = SU(2)_L \times  SU(3)_c \times U(1)_Y$, which is in contrast to other realizations which preserve a bigger $SU(2)_L \times SU(4) \times U(1)_A$ symmetry on the IR brane (see \cite{Lim:2007jv,Maru:2019lit,Maru:2019bjr}). %making it difficult to obtain the correct SM spectrum
The latter then leads to the problematic emergence of light exotic fermions from the large tensor representations for minimal models, a generic problem in GHGUT~\cite{Hosotani:2015hoa,Furui:2016owe,Hosotani:2016njs}, and obstructs the generation of masses for down-type quarks and/or charged leptons, as we will discuss further below.\footnote{Note that in orbifold constructions, the above symmetry breaking would be achieved with the help of brane scalars with large vacuum expectation values (vevs)~\cite{Angelescu:2021nbp,Hosotani:2008tx}.}

The corresponding BCs of the components of the $SU(6)$ gauge field~$A_\mu=A_\mu^a T^a$ read
\begin{equation}
\label{eq:bcs}
\begin{split}
A_\mu&= \left( \begin{array}{cc|ccc|c}
 (++) & (++) & (+-) & (+-) & (+-) & (--)\\
 (++) & (++) & (+-) & (+-) & (+-) & (--)\\
 \hline
 (+-) & (+-) & (++) & (++) & (++) & (--)\\
 (+-) & (+-) & (++) & (++) & (++) & (--)\\
 (+-) & (+-) & (++) & (++) & (++) & (--)\\
 \hline
 (--) & (--) & (--) & (--) & (--) & (--)\\
\end{array} \right),
\end{split}
\end{equation}
where $+(-)$ denote Neumann (Dirichlet) BCs on the (UV\ IR) branes, and the BCs for the scalars $A_5$ follow from flipping signs. 
The $(++)$ components feature massless zero modes, belonging to the unbroken gauge generators of $G_{\rm SM}$ in the 4D vector-boson ($A_\mu$) sector (upper left and central block) and to four EW-Higgs degrees of freedom in the 4D scalar sector ($A_5$), augmented with a $(\mathbf{3},\mathbf{1})_{-1/3}$ scalar leptoquark and a full scalar singlet (see last row, from left to right). We observe that all gauge bosons as well as the Higgs sector are contained in a {\it single} gauge field.

Before specifying the fermion content of the model, we note that from a composite Higgs perspective, the symmetry reduction on the branes, Eq.~\eqref{eq:breakingmod},
corresponds to a 4D CFT possessing a global $SU(6)$ symmetry, with the $SU(5)\supset G_{\rm{SM}}$ subgroup
weakly gauged, and $SU(6)$ being spontaneously broken to
$G_{\rm SM}$ in the infrared by condensation. The latter leads to the emergence of the ($A_5$) scalars, discussed above, as pNGBs. Explicit symmetry breaking will lift those to viable (and even potentially useful) mass regions, as we will see further below.

\subsection{Fermion embedding}

We introduce a minimal set of 5D fermions in $SU(6)$ representations, where the BCs (respecting the gauge symmetries on the respective branes) will lead to chiral SM-like zero modes that become massive via the Higgs mechanism.
In order to embed the full matter sector, one needs minimally a $\bf{20}$ and a $\bf{15}$ representation of $SU(6)$ for the up-type and down-type quarks, respectively \cite{Angelescu:2021nbp}, which have to be connected to form one light left handed (LH) quark doublet eigenstate interacting with both right handed (RH) quarks. This is realized by brane masses on the AdS boundaries. On the UV brane, these terms have to respect $SU(5)$, admitting only connections between the $SU(5)$ sub-representations of the bulk fields, which allows in principle to provide masses to all~the~SM~fermions.

Still, the UV interactions are too restricted to lead to a valid SM-fermion spectrum and therefore further boundary masses on the IR brane are added~\cite{Angelescu:2021nbp}, helping also to arrive at a viable one-loop Higgs potential. On top of the novel breaking pattern, it is the new boundary terms (that are now allowed from the symmetries) that lead to an admissible spectrum -- with, finite and different, down-type quark and charged-lepton masses.

The full SM spectrum, including massive neutrinos, is now reproduced from a minimal set of a $\bf{20}$, $\bf{15}$, $\bf{6}$ and a $\bf{1}$ bulk fermion per generation. Denoting the components of the
5D fields by the canonical symbols of
the SM-like zero modes they host, the decompositions for the LH modes read\\[4mm]
\begin{tabular}{cc}
\parbox{7cm}{$
 {\bf 20_L} \rightarrow {\bf ( 3,2)}_{1/6}^{-,+}  \oplus {\bf (3^*,1)}_{-2/3}^{-,+} \oplus {\bf (1,1)}_1^{-,+} 
 $\\[2mm]
 \hspace*{1.2cm}$ \oplus\,{\bf (3^*,2)}_{-1/6}^{-,+}  \oplus u_R {\bf (3,1)}_{2/3}^{-,-} \oplus {\bf (1,1)}_{-1}^{-,+} $
 }
& \quad
\parbox{7cm}{$ {\bf 15_L} \rightarrow  q_L {\bf (3,2)}_{1/6}^{+,+} 
        \oplus {\bf (3^*,1)}_{-2/3}^{+,-} \oplus e_R^c {\bf (1,1)}_1^{+,+}$\\[2mm]
         \hspace*{3cm}$\oplus {\bf (3,1)}_{-1/3}^{-,+}\oplus {\bf (1,2)}_{1/2}^{-,+}
$}\\[0.8cm]
\parbox{7cm}{$
        {\bf 6_L} \rightarrow d_R  {\bf (3,1)}_{-1/3}^{-,-} 
        \oplus l_L^c  {\bf (1,2)}_{1/2}^{-,-} \oplus \nu_R^c {\bf (1,1)}_0^{+,+}$}
&\hspace*{-3.8cm}
\parbox{11.2cm}{\begin{equation} {\bf 1_L} \rightarrow {\bf (1,1)}_{0}^{+,-}\,, \end{equation}
}
\end{tabular}\vspace{1.2mm}
while those for the RH modes are obtained flipping the BCs. We remark that the first (second) lines in the decompositions of the $\bf{20}$  and $\bf{15}$ correspond to a $\bf{10}$ ($\bf{10^*}$) and $\bf{10}$ ($\bf{5}$) of $SU(5)$, respectively.

It turns out that the modified breaking pattern \eqref{eq:breakingmod}
allows to lift the problematic exotic light states beyond LHC limits since the reduced SM-like symmetry on the IR brane admits a Neumannn BC for the $(\bf{1},\bf{1})_1^{-,+}$ in the $\bf{20}$, which would in a $SU(2)_L \times SU(4) \times U(1)_A$ symmetric theory have to be aligned with that of $u_R$ leading to a wrong-hypercharge RH electron-like zero mode \cite{Angelescu:2021nbp}. Note that, even though the $\bf{15}$ alone could in principle also host the $d_R$ and the lepton doublet, they would be mass-degenerate and that is why a $\bf{6}$ is introduced -- coupled to the $\bf{15}$ on the IR brane -- with the physical $d_R$ and $l_L^c$ finally residing mostly in the $\bf{6}$. The latter allows at the same time for a RH neutrino to generate (Dirac) neutrino masses which can be naturally light, i.e. $m_\nu\!<\!1$\,eV with ${\cal O}(1)$ input parameters, by mass-mixing it with the $\bf{1}$ on the IR brane (see \cite{Angelescu:2021nbp} for details).

The SM spectrum then follows straightforwardly after adding the mentioned UV and IR Lagrangians, connecting fields with the same $SU(5)$ and $G_{\rm SM}$ representations, respectively. Denoting LH (RH) spinors in representation ${\bf r}$ of $SU(6)$ and s of the unbroken group at the respective boundary by $\chi_{\alpha;{\bf r},{\rm s}}$ ($\bar \psi^{\dot{\alpha}}_{{\bf r},{\rm s}}$) and omitting flavor indices, the (most general) boundary terms read
\begin{equation} 
S_{UV} = \int \text{d}^4 x\big(
 M_u\, \psi_{{\bf{20}},10}\,\chi_{{\bf{15}},10}+\text{h.c.}\big)\, ,
\end{equation}  
\vspace{-6mm}
\begin{equation} 
S_{IR} \! = \!\!\int\!\!\!\text{d}^4 \!x \!\left(\!\frac{R}{R^\prime}\!\right)^{\!\!4}\!
        \big(M_{\tilde{u}} \psi_{{\bf{15}},(3^*\!\!,1)}\chi_{{\bf{20}},(3^*\!\!,1)} +\!M_d \chi_{{\bf{15}},(3,1)}\psi_{{\bf{6}},(3,1)} +\!M_{l} \chi_{{\bf{15}},(1,2)}\psi_{{\bf{6}},(1,2)} +\!M_{\nu} \chi_{{\bf{6}},1}\psi_{{\bf{1}}} +\!\text{h.c.}\big).\notag
\end{equation}
With these ingredients, our model can successfully accommodate the three massive SM-fermion generations and push all excitations above LHC limits. Here, the ${\bf (3^*,1)}_{-2/3}$ sector, linked via $M_u$ and $M_{\tilde{u}}$, is particularly interesting since for $M_{\tilde{u}}\!\neq\!0$ it contributes crucially to the Higgs potential. We close this section noting that every brane term is essential:  $M_u$ induces massive up-type quarks, $M_{\tilde{u}}$ helps in EW symmetry breaking (EWSB), $M_d, M_l$ lift the degeneracy between the down and charged lepton sectors, and $M_\nu$ allows for light neutrinos. The setup is summarized in Fig.~\ref{figures},~left~panel.

\section{Potential for the pNGBs}
\label{sec:CW}
The scalar potential depends on three real vevs: the Higgs vev $v$, the leptoquark vev $c$ and the singlet vev $s$. 
Using the Coleman-Weinberg one-loop formula~\cite{Falkowski:2006vi}, we obtain for the different contributions
\begin{equation}
\label{CW}
    V_r(v,c,s)=\frac{N_r}{(4\pi)^2}\int_0^\infty dp \, p^3\log(\rho_r(-p^2,v,c,s)),
\end{equation}
where $N_r=-4N_c$ for quarks, $N_r=3$ for gauge bosons, and $\rho_r$ are the spectral functions, with roots at $-p^2=m_{n;r}^2,\, n\in \mathbb{N}$ that encode the physical spectra. 

We start investigating the EWSB structure in the new warped $SU(6)$ framework by evaluating the potential along the Higgs direction ($s=c=0$). This is dominated by contributions from the top quark, the $W$- and $Z$-bosons, and from the mentioned third generation up-type exotic sector, stabilizing the potential, which leads to a limited set of free parameters~\cite{Angelescu:2021nbp} (neglecting the lighter generations of fermions as well as the bottom and tau sectors -- corresponding to moderately suppressed $M_{d,l}$). 
The EW sector of the potential thus depends on four parameters: $c_{15}$, $c_{20}$, $M_u$, and $M_{\tilde{u}}$, where $c_i \equiv m_i R$, with $m_i$ the Dirac bulk masses of the 5D fermions. We note that, fixing the IR scale at $1/R^{\prime} = 10$ TeV (evading collider constraints, see below), the correct $W$ boson mass is obtained for a dimensionful $SU(6)$ gauge coupling $g_5=g_*R^{1/2}\sim 3.8R^{1/2}$. 

\begin{figure}[t!]
	\hspace{-0.2cm}\includegraphics[scale=0.42]{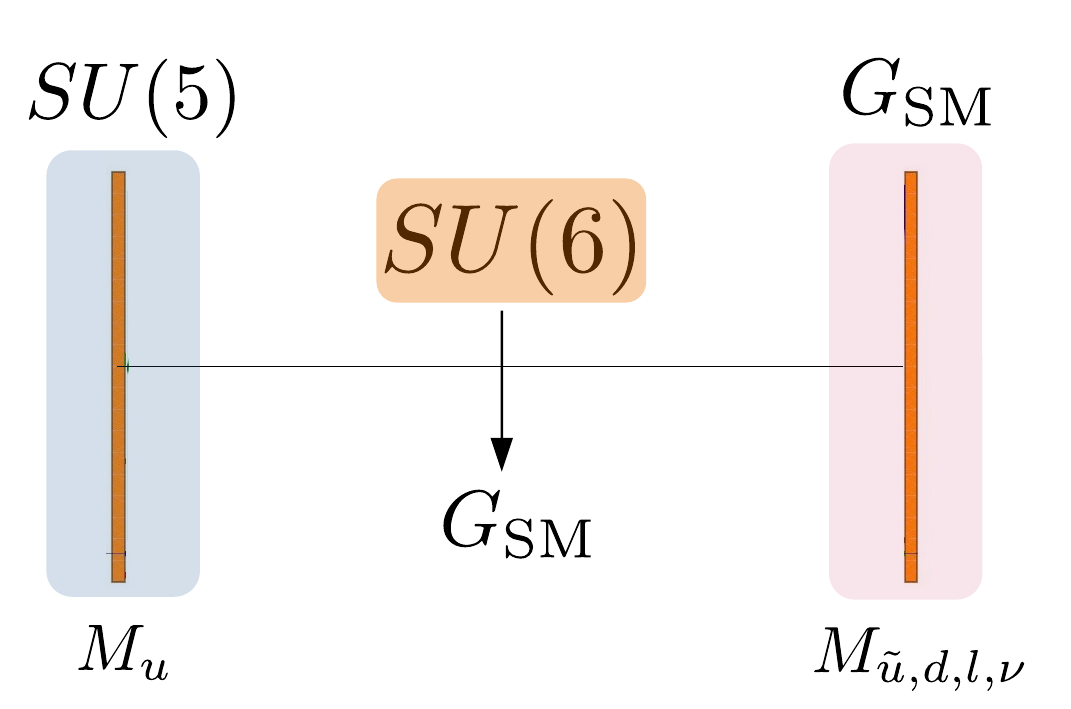}\quad
	\raisebox{-0.35cm}{\includegraphics[scale=0.58]{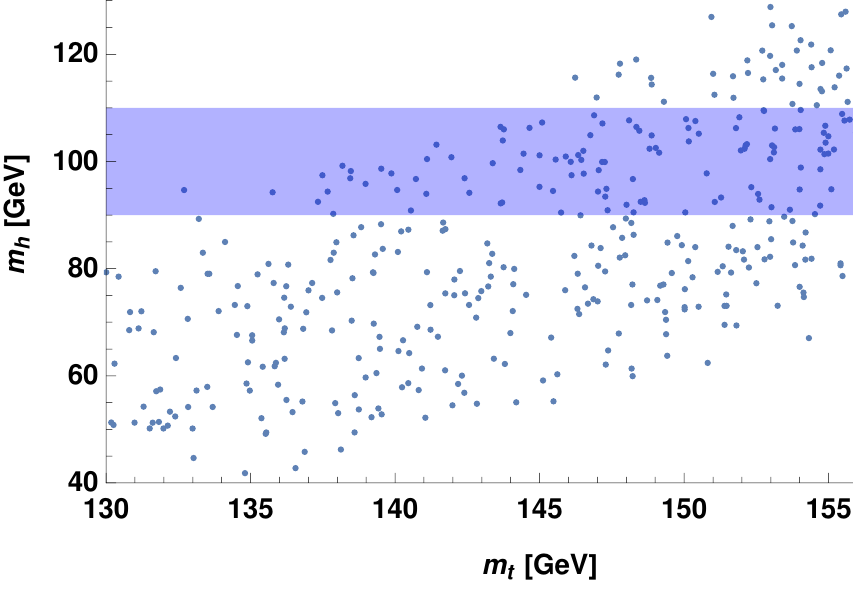}}\quad
	\raisebox{-0.35cm}{\includegraphics[scale=0.52]{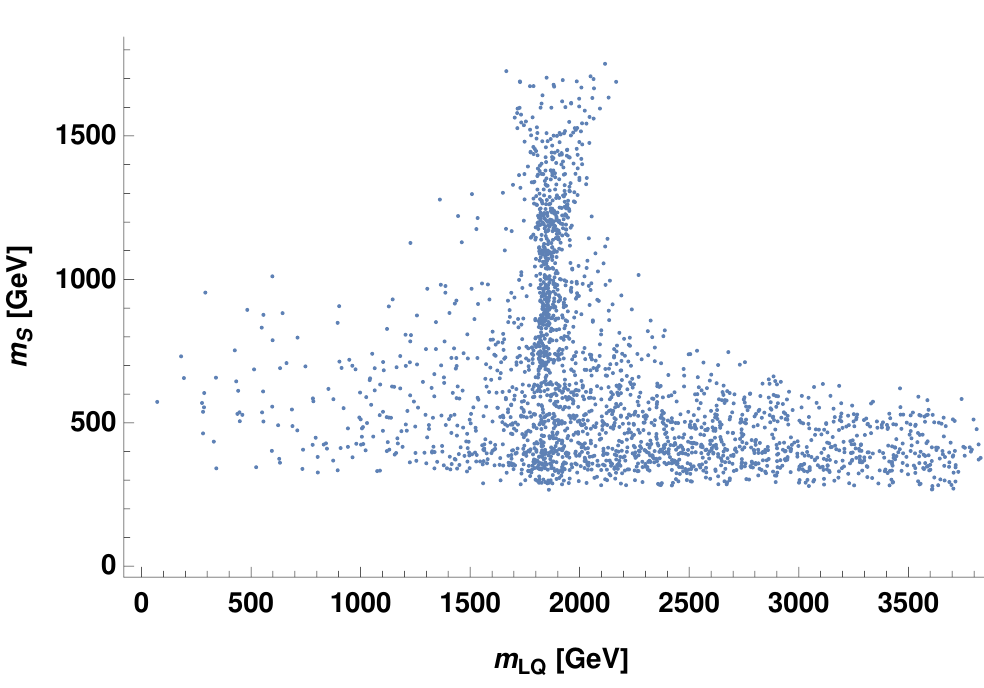}}
	\vspace{-0.3cm}
	\caption{Left: GHGUT setup. Middle: $m_h$ versus $m_t$ (at $\mu\!\sim\!f_\pi$), with the blue stripe highlighting the correct Higgs mass $m_h\in [90,110]$ \cite{Carmona:2014iwa}. Right: $m_{\textrm{S}}$ versus $m_{\textrm{LQ}}$. See text for details.}
	\label{figures}
\end{figure}

For our numerical study, we scan the third generation brane masses in  $0.1\!<\!M_{u,\tilde{u}}\!<\!3$,
with $c_i\!\sim\!{\cal O}(1)$, requiring $v\!\approx\! 246$\,GeV and a good fit to the fermion spectrum at $\mu \!\sim\!f_\pi$ (without~too light excitations), with $f_\pi\!=\!2/(g_\ast R^\prime)$ the pNGB decay constant. In the middle panel of Fig.~\ref{figures} we present the resulting $m_h(m_t)$. After fixing $m_t(\mu\!\!=\!\!f_\pi)\! \sim\! 140$\,GeV, the model predicts a light Higgs in excellent agreement with observation, which is a remarkable result due to the new setup~described~above~\cite{Angelescu:2021nbp}. 

Scanning further over the remaining parameters $c_1,c_6,M_d,M_l$, and $M_\nu$, with $0.1<M_{d,l,\nu}<3$, to reproduce the full fermion spectrum, we evaluate the potential \eqref{CW} along the leptoquark and singlet directions $c,s$, making sure that no vev is generated.
The resulting singlet mass $m_S$ versus $m_{\rm LQ}$ is plotted in the right panel of Fig.~\ref{figures}, exhibiting a broad range of viable masses, with prominent regions at $m_S \sim 500\,$GeV and $m_{\rm LQ}\sim 2\,$TeV. Interestingly, the triplet is in the right mass range to potentially explain the charged-current B-anomalies~\cite{Angelescu:2018tyl}. Moreover, the light pNGB singlet may develop a vev which could play a role in enhancing the first order phase transition and thereby allow for baryogenesis -- which is left for further investigation.

%%%%%%%%%%%%%%%%%%%%%%%%%%%%%%%%%%%%%%%%
\section{GHGUT Phenomenology}
\label{sec:pheno}

Since the hypercharge $U(1)_Y$ is contained in the upper-left $5\times 5$ block of \eqref{eq:bcs}, the EW gauge structure of $SU(6)$ GHGUT follows usual Georgi-Glashow $SU(5)$. % with a Weinberg angle of $\sin^2\theta_W=3/8$ which implies at the classical level $M_Z=\sqrt{8/5} M_W$.
Notable is the presence~of~light~$(+,-)$ vector bosons with
\begin{equation}
    m_{(+,-)}=\frac{2}{R^{\prime}\sqrt{2\log(\frac{R^{\prime}}{R})-1}}\sim 0.25/R^{\prime} < {R^\prime}^{-1}\,,
\end{equation}
corresponding to the $X, Y$ bosons from 4D GUTs.
%,in contrast to non-GUT $SO(5)\times U(1)$ GHU, which only contain exotic $(-,+)$ gauge bosons with larger $m_{(-,+)}\sim 2.4/R^{\prime}$.
However their much lower mass around the TeV scale opens up the exciting possibility of direct observation of these colored GUT states.  
Their profiles are similar to gauge boson zero-modes, featuring unsuppressed couplings to the first-generation,
\begin{equation}
\label{current}
    g_{\textrm{LQ}}\,({\cal X}_\mu^{\dagger})_i(y\, Q_L^i\gamma^{\mu}e^{c\,\dagger}_L  + y^\prime  \epsilon^{ij}(L^{c\,\dagger}_R)_j \gamma^\mu  d_R)/\sqrt{2}  +  \text{h.c.},
\end{equation}
where $({\cal X}_\mu^\dagger)_i = (Y_\mu,X_\mu)$, $\epsilon^{ij}$ is the antisymmetric  $SU(2)$ tensor, and $y^{(\prime)}\!\lesssim\!1$ parameterize the overlap between the gauge bosons and the fermionic zero modes in the extra dimension. In general this leads to tight constraints from non--resonant di-lepton searches (with leptoquarks in the t-channel)~\cite{Crivellin:2021egp}. While the exact computation of $g_{\textrm{LQ}}$ at low energies is beyond immediate scope, for $g_{\textrm{LQ}}\lesssim 1$ our benchmark with $m_{X,Y}\sim 0.25/R^\prime = 2.5$ TeV remains viable. 

It is interesting to note that, beyond collider limits, more aspects indicate a scale of ${R^\prime}^{-1}\!\sim\!10$\,TeV.
In fact, since $SU(6)$ does not feature a custodial symmetry, tree level contributions to the EW $T$-parameter point to a similar order of magnitude~\cite{Csaki:2002gy,Carena:2003fx,Casagrande:2008hr,Agashe:2013kxa} (which brings down the corrections to $\Delta T \approx 0.04$) and the same is true for expected flavor bounds~\cite{Csaki:2008zd}, which will be explored in detail in a future work. Turned around this means that the model could be discovered 'around the corner' in various branches, including flavor, precision tests, and collider searches, while Higgs-coupling measurements are less promising.
For the masses of the fermionic resonances, the setup predicts a rather large range from $0.1/R^\prime$ to $2.5/R^\prime$, including a light top-like exotic with $m_{\tilde T} \approx 0.3/R^\prime \approx 3$\,TeV -- a promising target for future collider searches. 

Finally, in generic GUTs the light X, Y bosons would mediate fast proton decay. However, in the 5D model at hand $q_L$ and $u_R$ reside in separate $SU(5)$ multiplets, which prohibits their dangerous $B\!-\!L$ conserving interactions with $X,Y$ inducing $p\rightarrow \pi_0 + e^+$. More generally, the model features a hidden baryon symmetry, which leads to $B$ conservation at each vertex (see~\cite{Angelescu:2021nbp}~for~details).

We close noting that different variations of the setup discussed above are possible, with similar broad features but important distinctions in particularities, realized by interchanging the boundary-breakings (and modifying the brane masses).
While detailed analyses will be presented elsewhere, we already comment on the impact on gauge-coupling unification: models with $H_0= SU(5)$ and $H_1=G_{\rm SM}$ feature an unbroken $SU(5)$ gauge group above the condensation scale $\sim {R^\prime}^{-1}$, where then unification has to be realized, requiring the presence of brane-kinetic terms \cite{Davoudiasl:2002ua,Carena:2002dz,Carena:2003fx}. On the other hand, the inverse choice, $H_0=G_{\rm SM}$ and $H_1= SU(5)$, induced by exchanging UV and IR BCs in \eqref{eq:bcs}, breaks the GUT group directly at the Planck scale $M_{\rm PL}$. This means it is not gauged in the first place from a 4D perspective and allows for high scale unification (still the theory features a single gauge coupling $g_5$ in the UV). Finally, realizing GUT breaking with a UV-brane scalar $\Phi_{\rm UV}$ that induces $H_0= SU(5) \to G_{\rm SM}$ at a large $M_{\rm GUT}$, admits a setup with 'conventional' unification with $ {R^\prime}^{-1} \ll M_{\rm GUT} < M_{\rm PL}$, relying on the logarithmic running in $AdS_5$ \cite{Agashe:2002pr,Pomarol:2000hp,Contino:2002kc,Randall:2001gb}.

%%%%%%%%%%%%%%%%%%%%%%%%%%%%%%%%%%%%%%%%
\section{Conclusions}
\label{sec:conc}
%%%%%%%%%%%%%%%%%%%%%%%%%%%%%%%%%%%%%%%%
We presented a minimal viable GHGUT in warped space based on a $SU(6)$ bulk symmetry, unifying the gauge interactions of the SM and their breaking sector in a simple gauge group \cite{Angelescu:2021nbp}. The full SM fermion spectrum can be naturally reproduced via appropriate embeddings for the bulk fields, avoiding the presence of ultra-light exotics. The new symmetry breaking pattern on the IR brane results in two additional pNGBs around the TeV scale, a leptoquark and a singlet, while the Higgs mass after EWSB is predicted at $m_h\sim 100$\,GeV. A global baryon number prohibits perturbative proton decay and a striking signature of the model are low-scale $X,Y$~vector~leptoquarks.

%%%%%%%%%%%%%%%%%%%%%%%%%%%%%%%%%%%%%%%%
\section*{Acknowledgments} 
%%%%%%%%%%%%%%%%%%%%%%%%%%%%%%%%%%%%%%%
\vspace{-0.3cm}
\small FG would like to thank the organizers of EPS-HEP2021 for the opportunity to present this work at the conference.
SB is supported by the “Excellence of Science - EOS” - be.h project n.30820817, and by the Strategic Research Program High-Energy Physics and the Research Council of the Vrije Universiteit Brussel.

\small
\bibliography{GHGUT}
\bibliographystyle{hunsrt}

\end{document}